\documentclass[11pt]{article}
\begin{document}
\title{Comment on "Is the Trineutron Resonance Lower in Energy than a Tetraneutron Resonance ?"}

\author{P. Tru\"ol\\ Physik-Institut, Universit\"at Z\"urich, CH-8090 Z\"urich, Switzerland\\
J.P. Miller\\ Department of Physics, Boston University, Boston, Massachusetts 02215, USA}
\maketitle
\newcounter{dummy}
In a recent letter Gandolfi {\em et al.}\cite{Gandolfi2017} reported their quantum Monte Carlo calculations for few-neutron systems. These seem to indicate that there exists a low-energy trineutron resonance and the authors suggest performing new experiments to search for it. The authors neglect to recognize that an experiment performed at the Los Alamos Meson Factory in the late seventies~\cite{Bistirlich1976,Miller1980} already showed that the trineutron system exhibits neither bound states nor low energy resonances. The experiment investigated radiative pion capture on a tritium target ($\pi^-$$^3$H$\rightarrow$nnn$\gamma$), a reaction in which the trineutron final state is observed free of any interference with another strongly interacting particle. Radiative pion capture on the hydrogen isotopes is theoretically well understood, as evidenced e.g. by a series of experiments with deuterium targets which provided the most precise information neutron-neutron scattering length and effective range~\cite{nn}. The photon spectrum in the tritium experiment is well described by a theoretical prediction~\cite{Phillips1974,Miller1980} reflecting the then state-of-the art three-nucleon calculations. The latter has also been applied to describe the unbound $nnp$ final state in radiative pion capture in $^3$He~\cite{Truoel1974}, where the photon spectrum also exhibits a strong transition to the $^3$H final state. The conceptually simple tritium experiment unfortunately also involved the difficult handling of a highly radioactive liquid tritium target, which presumably prohibits another attempt to repeat it.

\end{document}